\documentclass[12pt]{article}
\usepackage{amssymb,amsmath,epsfig}
\allowdisplaybreaks
\begin{document}
\title{\bf Charged Gravastars with Conformal Motion in $f(R,T)$ Gravity}

\author{M. Sharif \thanks{msharif.math@pu.edu.pk} and Arfa Waseem
\thanks{arfawaseem.pu@gmail.com}\\
Department of Mathematics, University of the Punjab,\\
Quaid-e-Azam Campus, Lahore-54590, Pakistan.}
\date{}

\maketitle

\begin{abstract}
This paper studies the effects of charge on a peculiar stellar
object, recognized as gravastar, under the influence of $f(R,T)$
gravity by considering the conjecture of Mazur and Mottola in
general relativity. The gravastar is also known as an alternative to
a black hole and is expressed by three distinct domains named as (i)
the interior domain, (ii) the intermediate shell and (iii) the
exterior domain. We analyze these domains for a specific $f(R,T)$
gravity model conceding the conformal Killing vectors. In the
interior domain, we assume that pressure is equal to negative energy
density which leads to the existence of repulsive force on the
spherical shell. The intermediate shell consists of
ultra-relativistic plasma and pressure which shows a direct relation
with energy density and counterbalances the repulsive force applied
by the interior domain. The exterior vacuum spherical domain is
taken to be the de Sitter spacetime illustrated by the
Reissner-Nordstr\"{o}m metric. We conclude that non-singular
solutions of charged gravastar with various physical properties such
as length, energy, entropy and equation of state parameter are
physically consistent.
\end{abstract}

\section{Introduction}

The composition of stellar structures and the phenomenon of
gravitational collapse are of significant interest in astrophysics
which have attracted many researchers since the development of
general relativity (GR). Gravitational collapse is responsible for
the creation of distinct massive stars such as neutron stars, white
dwarfs and black holes, referred to as compact objects. The ultimate
outcome of collapse entirely relies on the initial mass of the
celestial bodies. These remnants of collapse are broadly perceived
from both theoretical as well as observational perspectives. Mazur
and Mottola (2004) introduced a novel unique model, as an exact
solution of the Einstein field equations, in terms of dark and cold
compact object known as gravastar or the gravitational vacuum star.
Gravastar is described as the spherically symmetric highly compact
and singularity free object that can be anticipated as compact as
the black hole.

The structure of gravastar is expressed by three distinct domains in
which the interior domain is based on the de Sitter condensate state
and bounded by an extremely thin-shell composed of the
ultra-relativistic matter whereas the vacuum exterior is specified
by the Schwarzschild solution. The concept of gravastar is quite
fascinating for the researchers as it can solve the two basic
problems related to black holes, i.e., the information paradox and
singularity problem. Despite several observational and theoretical
accomplishments, there are still various challenging issues that
stimulate the researchers to examine different alternatives in which
the outcomes of collapse are giant stars without event horizons.
Some examples of such stellar objects are bose superfluids, black
stars, dark energy stars and gravastars that possess no event
horizon and singularity (Chapline et al. 2003; Lobo 2006; Chirenti
and Rezzolla 2008). Among these objects, gravastars have motivated
the researchers to discuss their structures using different
approaches.

In the framework of GR, Visser and Wiltshire (2004) discussed the
stability of gravastar and found that there are various equations of
state (EoS) that lead to dynamically stable distribution of
gravastar. Cattoen et al. (2005) presented a generalized class of
gravastars that possesses anisotropic pressure without existence of
very small thin-shell. Carter (2005) obtained new spherical
solutions of gravastar and observed distinct qualitative behavior of
EoS on the interior and exterior domains. Bili\'{c} et al. (2006)
analyzed the interior solutions of gravastar by replacing the de
Sitter metric with Born-Infeld phantom and determined that their
solutions can describe the dark compact objects at the center of
galaxies. Horvat and Iliji\'{c} (2007) investigated energy
conditions at the shell of gravastar and inspected the stable
structure against radial perturbations via speed of sound on the
shell. Several researchers (Broderick and Narayan 2007; Chirenti and
Rezzolla 2007; Rocha et al. 2008; Cardoso et al. 2008; Harko et al.
2009; Pani et al. 2009) studied interior solutions using various EoS
corresponding to different conjectures.

The electromagnetic field has significant role in the study of
structure evolution as well as stability of collapsing objects. The
electromagnetic forces, i.e., Coulomb and magnetic, are the crucial
sources to generate the repulsive effects which counterbalance the
attractive gravitational force. In order to keep a stellar object in
its equilibrium state, a star requires an immense amount of charge
to overcome the strong gravitational pull. In the study of
gravastar, Lobo and Arellano (2007) constructed different gravastar
models along with nonlinear electrodynamics and analyzed their
solutions as well as some specific structural features. Horvat et
al. (2009) extended the concept of gravastar by including the
electrically charged component and evaluated the EoS parameter,
surface redshift and speed of sound for the interior and exterior
domains. On the same ground, Turimov et al. (2009) explored the
dipolar magnetic field distribution and determined solutions of the
Maxwell equations for the internal domain of a slowly rotating
gravastar.

It is well-known that GR has achieved marvelous results in resolving
the various hidden puzzles of the cosmos. However, some
observational facts about current accelerating universe and dark
matter inflicted a theoretical barrier to this theory (Riess et al.
1998; Perlmutter et al. 1999; de Bernardis et al. 2000; Peebles and
Ratra 2003). In order to alleviate these problems, modified or
alternative theories to GR are identified as the most fruitful
perspectives. The $f(R)$ theory of gravity (Capozziello 2002) is
recognized as the direct generalization of GR which is developed by
placing a generic function $f(R)$ instead of curvature scalar ($R$)
in the Einstein-Hilbert action. The interesting concept of the
coupling between matter and geometry leads to different modified
theories such as $f(R,T)$ gravity (Harko et al. 2011), $f(R,T,
R_{\gamma\delta}T^{\gamma\delta})$ theory (Haghani et al. 2013;
Odintsov and S\'{a}ez-G\'{o}mes 2013) and $f(\mathcal{G},T)$ gravity
(Sharif and Ikram 2016), where $T$ exhibits the trace of
energy-momentum tensor (EMT) and $\mathcal{G}$ stands for
Gauss-Bonnet invariant. Among these modified theories, $f(R,T)$
gravity received much attention by several researchers in describing
various astrophysical as well as cosmological scenarios with and
without electric field (Houndjo 2012; Sharif and Zubair 2012, 2013a,
b; Shabani and Farhoudi 2013, 2014; Sharif and Zubair 2014; Moraes
2015; Moraes et al. 2016; Sharif and Siddiqa 2017; Das et al. 2017;
Sharif and Waseem 2018a, b; Deb et al. 2018a, b; Sharif and Siddiqa
2018, 2019; Sharif and Nawazish 2019; Sharif and Waseem 2019).

In order to obtain a natural systematic relationship between matter
and geometry ingredients for giant compact stars through the field
equations, an inheritance symmetry that contains a set of conformal
Killing vectors plays a fundamental role. These vectors are utilized
to attain exact analytic solutions of the field equations in more
appropriate expressions in comparison with other analytical
approaches. Using these vectors, the highly nonlinear partial
differential equations can be transformed into a system of ordinary
differential equations. Usmani et al. (2011) employed these Killing
vectors to study the geometry of charged gravastar and determined
solutions for different domains of charged gravastar in terms of
conformal vector. Recently, the study of gravastar has attracted
many people in modified theories of gravity. Das et al. (2017)
discussed the conjecture of gravastar in $f(R,T)$ framework and
analyzed its physical characteristics graphically corresponding to
different EoS. Shamir and Ahmad (2018) presented non-singular
solutions of gravastar and evaluated mathematical expressions of
various physical parameters in $f(\mathcal{G},T)$ scenario.

In this paper, we discuss the geometry of gravastar which is
internally charged in $f(R,T)$ gravity conceding the conformal
motion. For $R+2\beta T$ gravity model, we analyze the structure of
charged gavastar described by three distinct domains and observe
some feasible features graphically. The paper is displayed according
to the following pattern. Next section deals with the mathematical
fundamentals of $f(R,T)$ gravity. Section \textbf{3} demonstrates
solutions of the field equations in terms of conformal vector. In
section \textbf{4}, we discuss the structure of charged gravastar
corresponding to different EoS. Section \textbf{5} provides
interesting physical aspects of charged gravastar shell like proper
length, EoS parameter, entropy and energy. The concluding remarks
are presented in the last section.

\section{Fundamentals of $f(R,T)$ Gravity}

The action of $f(R,T)$ theory associated with matter Lagrangian
($L_{m}$) and electric field is specified by (Harko et al. 2011)
\begin{equation}\label{1}
\mathcal{I}_{f(R,T)}=\int \left[\frac{f(R,
T)}{2\kappa}+L_{m}+L_{e}\right]\sqrt{-g}d^{4}x,
\end{equation}
with $\kappa=1$ being a coupling constant, $g$ exhibits determinant
of $g_{\gamma\delta}$, $L_{e}=nF_{\gamma\delta}F^{\gamma\delta}$,
$n$ represents an arbitrary constant, $F_{\gamma\delta}
=\vartheta_{\delta,\gamma} -\vartheta_{\gamma,\delta}$ expresses the
electromagnetic field tensor and $\vartheta_{\gamma}$ symbolizes the
four-potential. The corresponding field equations are
\begin{eqnarray}\nonumber
f_{R}(R,T)R_{\gamma\delta}&-&(\nabla_{\gamma}\nabla_{\delta}
-g_{\gamma\delta}\Box)f_{R}(R,T)-\frac{1}{2}g_{\gamma\delta}f(R,T)
\\\label{2}&=&T_{\gamma\delta}+E_{\gamma\delta}-(\Theta_{\gamma\delta}
+T_{\gamma\delta})f_{T}(R,T),
\end{eqnarray}
where $f_{R}(R,T)=\frac{\partial f(R,T)}{\partial
R},~f_{T}(R,T)=\frac{\partial f(R,T)}{\partial T}$,
$\Box=g^{\gamma\delta}\nabla_{\gamma}\nabla_{\delta}$,
$\nabla_{\gamma}$ reveals the covariant derivative,
$\Theta_{\gamma\delta}$ is characterized by
\begin{equation}\label{3}
\Theta_{\gamma\delta}=g^{\mu\upsilon}\frac{\delta
T_{\mu\upsilon}}{\delta g^{\gamma\delta}}=
g_{\gamma\delta}L_{m}-2T_{\gamma\delta}-
2g^{\mu\upsilon}\frac{\partial^{2}L_{m}}{\partial
g^{\gamma\delta}\partial g^{\mu\upsilon}},
\end{equation}
and electromagnetic EMT is evaluated as
\begin{equation}\label{4}
E_{\gamma\delta}=\frac{1}{4\pi}\left(\frac{F^{\mu\upsilon}
F_{\mu\upsilon}g_{\gamma\delta}}{4}-F^{\mu}_{\gamma}F_{\delta\mu}\right).
\end{equation}
The covariant divergence of Eq.(\ref{2}) yields (Harko et al. 2011)
\begin{equation}\label{5}
\nabla^{\gamma}T_{\gamma\delta}=\frac{f_{T}}{1-f_{T}}
\left[(T_{\gamma\delta}+\Theta_{\gamma\delta})\nabla^{\gamma}(\ln
f_{T})-\nabla^{\gamma}(\frac{g_{\gamma\delta}T}{2}-
\Theta_{\gamma\delta})-\frac{\nabla^{\gamma}E_{\gamma\delta}}{f_{T}}\right].
\end{equation}
This equation manifests that the EMT does not obey the conservation
law in such modified theories unlike GR. In order to discuss the
cosmological as well as astrophysical scenarios, the distribution of
matter plays a pivotal role. All the non-vanishing constituents of
EMT express the dynamical quantities along with various physical
influences. To discuss the structure of charged gravastar, we
consider perfect fluid matter distribution given as
\begin{equation}\label{6}
T_{\gamma\delta}=(\rho+p)U_{\gamma}U_{\delta}-pg_{\gamma\delta},
\end{equation}
where $\rho$ indicates matter energy density, $p$ stands for
pressure and $U_{\gamma}$ acts as four velocity in comoving
coordinates satisfying $U^{\gamma}U_{\gamma}=1$. In matter
distribution, there are different choices of $L_{m}$. Here, we
assume $L_{m}=-p$ which provides $\frac{\partial^{2}L_{m}}{\partial
g^{\gamma\delta}\partial g^{\mu\upsilon}}=0$ (Harko et al. 2011) and
$\Theta_{\xi\eta}=-2T_{\gamma\delta}-p g_{\gamma\delta}$.

In the analysis of compact objects, it is observed that the
celestial configurations presently exist in nonlinear regime. For
their proper structure evolution, one must need to inspect their
linear behavior and hence, we choose an independent linear model as
\begin{equation}\label{7}
f(R,T)=R+2\beta T,
\end{equation}
where $T=\rho-3p$ and $\beta$ is a coupling parameter. In this
theory, the addition of $T$ provides more modified forms of GR in
comparison with $f(R)$ gravity. This functional form is
astronomically applicable to discuss different cosmological issues
and expresses the Lambda cold dark matter ($\Lambda$CDM) model in
$f(R,T)$ gravity (Zubair et al. 2016). This model was firstly
proposed by Harko et al. (2011) and has broadly been employed to
study the features of different astrophysical objects. Recently, Das
et al. (2017) used this model to analyze exact non-singular
solutions of collapsing object and presented physically consistent
properties of gravastar. Inserting the above model along with the
considered choice of $L_{m}$ in Eq.(\ref{2}), we obtain
\begin{equation}\label{8}
G_{\gamma\delta}=T_{\gamma\delta}+E_{\gamma\delta}+\beta
Tg_{\gamma\delta}+2\beta(T_{\gamma\delta}+pg_{\gamma\delta}),
\end{equation}
where $G_{\gamma\delta}$ shows the standard Einstein tensor. The
covariant divergence (\ref{5}) corresponding to $R+2\beta T$ model
becomes
\begin{equation}\label{9}
\nabla^{\gamma}T_{\gamma\delta}=\frac{-1}{1+2\beta}\left[\beta
g_{\gamma\delta}\nabla^{\gamma}T+2\beta\nabla^{\gamma}
(pg_{\gamma\delta})+\nabla^{\gamma}E_{\gamma\delta}\right].
\end{equation}
For $\beta=0$, the original conserved results of GR in the presence
of electric field can be retrieved.

\section{Field Equations With Conformal Motion}

In order to characterize the interior of gravastar, we consider
static spherically symmetric spacetime as
\begin{equation}\label{10}
ds^{2}_{-}=e^{\lambda(r)}dt^{2}-e^{\chi(r)}dr^{2}
-r^{2}d\theta^{2}-r^{2}\sin^{2}\theta d\phi^{2}.
\end{equation}
This metric along with Eqs.(\ref{6}) and (\ref{8}) leads to the
following field equations
\begin{eqnarray}\label{11}
\frac{1}{r^{2}}-e^{-\chi}\left(\frac{1}{r^{2}}-\frac{\chi'}{r}\right)&=&\rho
+\beta(3\rho-p)+\frac{q^{2}}{8\pi r^{4}},
\\\label{12}
e^{-\chi}\left(\frac{1}{r^{2}}+\frac{\lambda'}{r}\right)-\frac{1}{r^{2}}&=&p
-\beta(\rho-3p)-\frac{q^{2}}{8\pi r^{4}},
\\\label{13}
e^{-\chi}\left(\frac{\lambda''}{2}-\frac{\chi'}{2r}+\frac{\lambda'}{2r}
+\frac{\lambda'^{2}}{4}-\frac{\chi'\lambda'}{4}\right)&=&p-\beta
(\rho-3p)+\frac{q^{2}}{8\pi r^{4}},
\end{eqnarray}
where prime reveals differentiation corresponding to $r$ and $q$
indicates the spherical charge defined as
\begin{equation}\label{14}
q(r)=4\pi\int_{0}^{r}\xi(r)r^{2}e^{\chi/2}dr,\quad\quad
E=\frac{q}{4\pi r^{2}}.
\end{equation}
Here $\xi$ and $E$ denote the surface charge density and electric
field intensity, respectively. The covariant divergence given by
Eq.(\ref{9}) provides (Moraes et al. 2016)
\begin{equation}\label{15}
p'+\frac{\lambda'}{2}(\rho+p)=\frac{-1}{1+2\beta}\left\{\beta(\rho'-p')
-\frac{qq'}{4\pi r^{4}}\right\}.
\end{equation}

In the analysis of astrophysical objects, for a natural relationship
between geometry and matter ingredients described by the field
equations, we assume a renowned inheritance symmetry which contains
a set of conformal Killing vectors expressed by the relation
\begin{equation}\label{16}
\mathfrak{L}_{\zeta}g_{\gamma\delta}=g_{\eta\delta}\zeta^{\eta}_{~;\gamma}
+g_{\gamma\eta}\zeta^{\eta}_{~;\delta}=\varphi(r)g_{\gamma\delta},
\end{equation}
where $\mathfrak{L}$ acts as a Lie derivative operator and
$\varphi(r)$ denotes the conformal vector. Using Eq.(\ref{10}) in
(\ref{16}), it follows that (Rahaman et al. 2014)
\begin{equation}\nonumber
\zeta^{1}\lambda'=\varphi,\quad \zeta^{1}=\frac{r\varphi}{2},\quad
\zeta^{1}\chi'+2\zeta^{1}_{,1}=\varphi,
\end{equation}
yielding
\begin{eqnarray}\label{17}
e^{\lambda}=a^{2}r^{2},\quad
e^{\chi}=\left(\frac{b}{\varphi}\right)^{2},
\end{eqnarray}
with $a$ and $b$ as the integration constants. Inserting these
solutions in Eqs.(\ref{11})-(\ref{13}), we obtain
\begin{eqnarray}\label{19}
\frac{1}{r^{2}}\left(1-\frac{\varphi^{2}}{b^{2}}\right)
-\frac{2\varphi\varphi'}{b^{2}r}&=&\rho+\beta(3\rho-p)
+\frac{q^{2}}{8\pi r^{4}},
\\\label{20}
\frac{3\varphi^{2}}{b^{2}r^{2}}-\frac{1}{r^{2}}&=&p
-\beta(\rho-3p)-\frac{q^{2}}{8\pi r^{4}},
\\\label{21}
\frac{\varphi^{2}}{b^{2}r^{2}}+\frac{2\varphi\varphi'}
{b^{2}r}&=&p-\beta(\rho-3p)+\frac{q^{2}}{8\pi r^{4}}.
\end{eqnarray}
These lead to the explicit forms of density, pressure and electric
field as follows
\begin{eqnarray}\label{22}
\rho&=&\frac{b^{2}(1+2\beta)+4\beta\varphi^{2}-2r\varphi\varphi'(3+8\beta)}
{2b^{2}r^{2}(1+6\beta+8\beta^{2})},\\\label{23}
p&=&-\frac{b^{2}(1+2\beta)+4\varphi^{2}(1+3\beta)+2r\varphi\varphi'}
{2b^{2}r^{2}(1+6\beta+8\beta^{2})},\\\label{24}2\pi E^{2}&=&
\frac{q^{2}}{8\pi r^{4}}=\frac{1}{2}\left[\frac{1}{r^{2}}\left(1
-\frac{2\varphi^{2}}{b^{2}}\right)+\frac{2\varphi\varphi'}{b^{2}r}\right].
\end{eqnarray}
We note that the electric field is independent of $f(R,T)$ model
parameter $\beta$ and expresses the same form as in GR. For
$\beta=0$, all these equations reduce to the field equations of GR
admitting conformal Killing vectors (Usmani et al. 2011).

\section{Solutions of Gravastar in $f(R,T)$ Gravity}

This section is devoted to discuss the geometrical description of
the constructed solutions of gravastar in $f(R,T)$ gravity
corresponding to (i) the interior domain, (ii) the intermediate
shell and (iii) the exterior domain. The interior geometry of a star
is surrounded by an extremely thin-shell which is comprised of some
ultra-relativistic fluid. On the other hand, the exterior domain is
purely vacuum and for charged gravastar, the Reissner-Nordstr\"{o}m
metric can be regarded feasible for this outer region. Since the
shell is considered to be immensely thin possessing a finite width
lying in the range $r_{1}=\mathcal{R}<r<\mathcal{R}+\epsilon=r_{2}$,
where $r_{1}$ and $r_{2}$ exhibit the inner and outer radii of
charged gravastar under consideration, respectively and
$r_{2}-r_{1}=\epsilon$ reveals thickness of the shell. Therefore,
the structure of charged gravastar can be characterized by three
domains depending upon the EoS as follows
\begin{itemize}
\item Interior domain $(\mathcal{D}_{1})$ $\Rightarrow$ $0\leq
r<r_{1}$ with $p+\rho=0$,
\item Intermediate shell $(\mathcal{D}_{2})$ $\Rightarrow$ $r_{1}\leq
r\leq r_{2}$ with $p=\rho$,
\item Exterior domain $(\mathcal{D}_{3})$ $\Rightarrow$ $r_{2}<r$
with $p=\rho=0$.
\end{itemize}

\subsection{The Interior Domain}

In this region, we assume the EoS as $\omega=\frac{p}{\rho}$, where
$\omega$ represents the EoS parameter and for $\omega=-1$, we have
$p=-\rho$ which expresses the EoS for dark energy. From
Eqs.(\ref{22}) and (\ref{23}), we obtain a connection between the
matter variables and metric functions in the following form
\begin{equation}\label{25}
\rho+p=\frac{2\varphi(1+4\beta)(\varphi-r\varphi')}{b^{2}r^{2}(1
+2\beta)(1+4\beta)}=\frac{2\varphi(\varphi-r\varphi')}{b^{2}r^{2}(1
+2\beta)}.
\end{equation}
Employing the ansatz $\rho+p=0$ in the above equation, it follows
that either $\varphi=0$, or $\varphi-r\varphi'=0$. We consider
$\varphi-r\varphi'=0$ which provides a solution of conformal vector
as $\varphi=\varphi_{0}r$, where $\varphi_{0}$ is a dimensionless
integration constant. This solution of conformal vector yields the
following analytic expressions of all the physical parameters
\begin{eqnarray}\nonumber
\rho&=&\frac{1+2\beta-6r^{2}\tilde{\varphi}_{0}^{2}-12\beta
r^{2}\tilde{\varphi}_{0}^{2}}{2r^{2}(1+6\beta+8\beta^{2})}=\frac{(1
+2\beta)(1-6r^{2}\tilde{\varphi}_{0}^{2})}{2r^{2}(1+2\beta)(1+4\beta)}
\\\label{26}&=&\frac{1-6r^{2}\tilde{\varphi}_{0}^{2}}{2r^{2}(1+4\beta)}
=-p,\\\label{27}
e^{\lambda}&=&e^{-\chi}=\tilde{\varphi}_{0}^{2}r^{2},\\\label{28}
E^{2}&=& \frac{q^{2}}{16\pi^{2} r^{4}}=\frac{1}{4\pi
r^{2}},\\\label{29} \xi&=&\frac{\tilde{\varphi}_{0}}{r\sqrt{4\pi}},
\end{eqnarray}
with $\tilde{\varphi}_{0}=\frac{\varphi_{0}}{b}$ as a constant
having dimension inverse of $r$.

Equation (\ref{26}) implies that
$\frac{1-6r^{2}\tilde{\varphi}_{0}^{2}}{2r^{2}(1+4\beta)}>0$ which
further yields the constraints either
$\tilde{\varphi}_{0}^{2}<\frac{1}{6r^{2}}$ and $\beta>\frac{-1}{4}$,
or $\tilde{\varphi}_{0}^{2}>\frac{1}{6r^{2}}$ and
$\beta<\frac{-1}{4}$. Here, we consider the former constraints. The
positive behavior of these constraints describe the positive matter
density and negative pressure which corresponds to an outward
directed force exerted by the interior domain and shows consistency
with the physical interpretation of charged gravastar. If we suppose
that $\frac{1-6r^{2}\tilde{\varphi}_{0}^{2}}{2r^{2}(1+4\beta)}<0$,
then it leads to a collapsing phenomenon with negative energy
density and positive pressure which is not considered here. Hence
for the physically viable description of a charged gravastar, we
assume that our solutions satisfy the constraint
$0<\tilde{\varphi}_{0}^{2}<\frac{1}{6r^{2}}$.

Equation (\ref{27}) describes a direct relation of the metric
potentials with the radius of charged gravastar. The electric field
and charge density are inversely proportional to $r$ as expressed in
Eqs.(\ref{28}) and (\ref{29}), respectively. Based on Eq.(\ref{11}),
the gravitational mass ($\mathcal{M}$) in the interior of charged
gravastar can be given as
\begin{equation}\label{30}
\mathcal{M}(\mathcal{R})=\frac{1}{2}\int^{r_{1}=\mathcal{R}}_{0}
\left(\rho+\beta(3\rho-p)+2\pi E^{2}\right)r^{2}dr.
\end{equation}
With the help of Eqs.(\ref{26}) and (\ref{28}), the solution of
Eq.(\ref{30}) leads to the following form of gravitational mass
\begin{equation}\label{31}
\mathcal{M}(\mathcal{R})=\frac{\mathcal{R}}{2}\left(1
-\tilde{\varphi}_{0}^{2}\mathcal{R}^{2}\right).
\end{equation}
This suggests that the gravitational mass vanishes at
$\mathcal{R}=0$ and exhibits a positive as well as regular behavior
due to the constraint $\tilde{\varphi}_{0}^{2}<\frac{1}{6r^{2}}$.
Hence, the gravitational mass does not yield a singularity. The
graphical behavior of mass-radius direct relation is presented in
Figure \textbf{1} which indicates physically viable characteristic
of compact stellar objects.
\begin{figure}\center
\epsfig{file=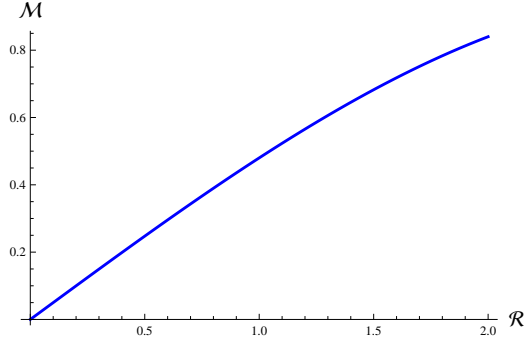,width=0.5\linewidth}\\
\caption{Behavior of $\mathcal{M}$ ($M_{\odot}$) versus
$\mathcal{R}$ ($km$) of charged gravastar with
$\tilde{\varphi}_{0}<\frac{1}{\sqrt{6}\mathcal{R}}$.}
\end{figure}

It is worthwhile to mention here that the mathematical description
as well as physics of the interior domain coincide with the
electromagnetic mass model (Ray et al. 2008). The significant reason
behind this is the inclusion of electric field in the geometry of
gravastar. As a result of spherical collapse, this mass produces the
attractive force which counterbalances the repulsive force generated
by the electromagnetic field. Also, in the framework of accelerating
cosmos, the EoS $\rho=-p$ indicates a repulsive pressure which may
correspond to dark energy, an enigmatic force responsible for the
current phase of the inflation. Thus we can say that charged
gravastar provides a connection with the dark star (Chan et al.
2009a, b).

\subsection{The Intermediate Shell of Charged Gravastar}

Here, we consider that the non-vacuum shell of charged gravastar is
formulated by ultra-relativistic fluid satisfying the EoS $\rho=p$.
In association with the cold baryonic cosmos, Zel'dovich (1972)
developed the concept of such type of fluid recognized as stiff
fluid. But in the present framework, it may be due to some thermal
oscillations with zero chemical potential or due to the
gravitational quantities at negligible temperature (Mazur and
Mottola 2004). Several researchers (Carr 1975; Wesson 1978; Madsen
et al. 1992; Braje and Romani 2002; Linares et al. 2004) have widely
utilized this fluid to discuss different cosmological as well as
astrophysical scenarios.

It is observed that finding exact solutions of the field equations
for a non-vacuum shell region is a difficult task in the presence of
stiff fluid. However, such solutions can be obtained within the
range suggested for the thin-shell, i.e., $0<e^{-\chi}\ll1$
admitting the conformal motion. Implementing this range along with
stiff fluid EoS in Eqs.(\ref{22}) and (\ref{23}), we evaluate
\begin{equation}\label{32}
(1+2\beta)(b^{2}-2\varphi^{2}-4r\varphi\varphi')=0.
\end{equation}
Again, we have two choices that either $1+2\beta=0$, or
$b^{2}-2\varphi^{2}-4r\varphi\varphi'=0$. The former condition
provides $\beta=-\frac{1}{2}$ and we neglect this choice to avoid
singular solution. The later condition leads to
\begin{equation}\label{33}
\varphi^{2}=\frac{b^{2}}{2}-\frac{\varphi_{1}}{r},
\end{equation}
with $\varphi_{1}>0$ being the integration constant. On substituting
this solution, Eqs.(\ref{22})-(\ref{24}) become
\begin{eqnarray}\label{34}
\rho&=&\frac{1}{2r^{2}(1+2\beta)}\left(1-\frac{3\tilde{\varphi}_{1}}{r}\right)
=p,\\\label{35}e^{\lambda}&=&a^{2}r^{2},\\\label{36}
e^{-\chi}&=&\frac{1}{2}-\frac{\tilde{\varphi}_{1}}{r},\\\label{37}
E^{2}&=&\frac{q^{2}}{16\pi^{2}r^{4}}=\frac{3\tilde{\varphi}_{1}}{4\pi
r^{3}},\\\label{38}
\xi&=&\frac{\sqrt{3}\tilde{\varphi}_{1}}{4\sqrt{2\pi}r^{3}}
\sqrt{\frac{r}{\tilde{\varphi}_{1}}-2},
\end{eqnarray}
here $\tilde{\varphi}_{1}=\frac{\varphi_{1}}{b^{2}}$ has the
dimension of $r$.

In this region, electric field depends on $\tilde{\varphi}_{1}$ and
shows an inverse relation with the radius. The charge density
demonstrates that the real value of $\xi$ can be found only for the
constraint $\tilde{\varphi}_{1}<\frac{r}{2}$. Combining this
condition with $\varphi_{1}>0$, we observe that the constructed
solutions for the shell region of charged gravastar are consistent
only for the range $0<\tilde{\varphi}_{1}<\frac{r}{2}$. The
interpretation of energy density in Eq.(\ref{34}) provides that the
EoS $\rho=p=0$ for the de-Sitter exterior region leads to
$\tilde{\varphi}_{1}=\frac{r}{3}$, which obviously satisfies the
constraint $\tilde{\varphi}_{1}<\frac{r}{2}$. Since, in the shell
domain $\mathcal{D}_{2}$, we have $\mathcal{R}\leq r\leq
\mathcal{R}+\epsilon$, therefore, from the assumption
$e^{-\chi}\ll1$, we obtain $\epsilon\ll1$.
\begin{figure}\center
\epsfig{file=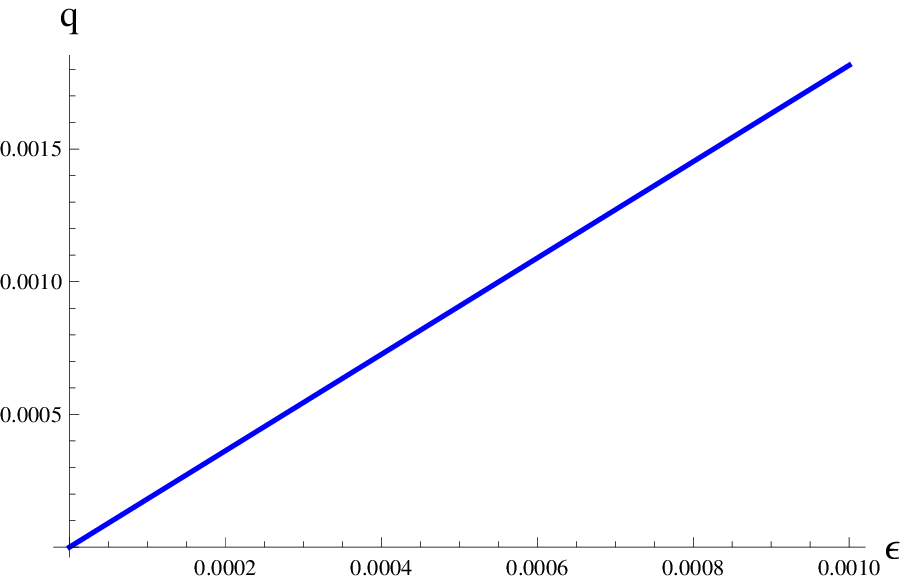,width=0.48\linewidth}
\epsfig{file=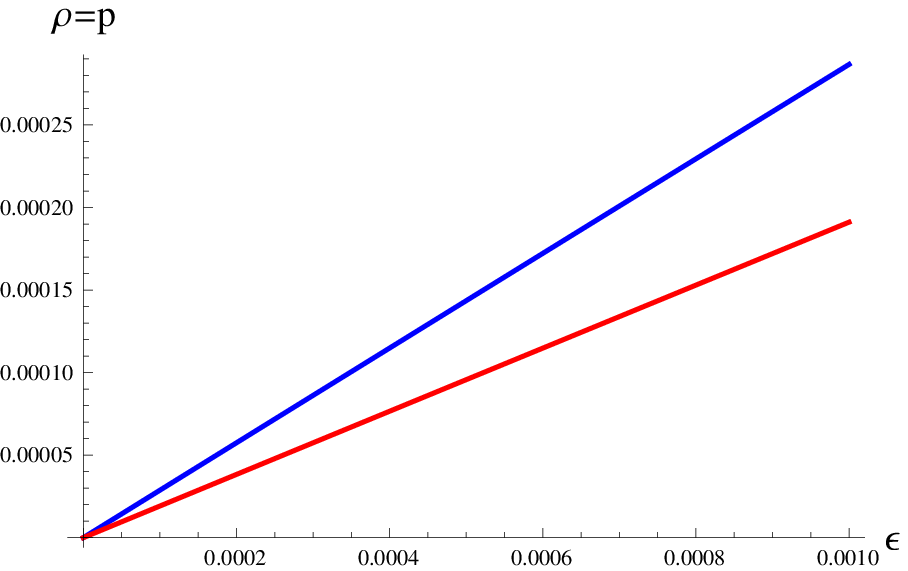,width=0.48\linewidth}\\
\caption{Behavior of $q$ (left) and $\rho=p$ (right) for
$\beta=\frac{1}{2}$ (blue) and $\beta=1$ (red) versus thickness of
the shell of charged gravastar with
$\tilde{\varphi}_{1}<\frac{r}{2}$.}
\end{figure}

The graphical behavior of charge $(q)$ and $\rho=p$ with respect to
thickness is displayed in Figure \textbf{2} which exhibits that the
ultra-relativistic fluid as well as charge show more effects towards
the outer domain as compared to the inner border of the shell. In
literature, it  is found that the value of model parameter $\beta$
lies in the interval $[-1,1]$ to discuss the structural properties
of self-gravitating objects. But in the present context, we have
observed that $\beta=-\frac{1}{2}$ yields singularity in the domain
$\mathcal{D}_{2}$. Due to this reason, we neglect the negative
values of $\beta$ and show the behavior of energy density for two
different positive values of $\beta$. The graphical analysis of
energy density shows that charged gravastars possess denser
structure for small values of $\beta$ and their denseness decreases
with increasing value of the model parameter.

\subsection{The Exterior Domain and Matching Constraints}

Finally, we discuss the vacuum region $\mathcal{D}_{3}$, when the
exterior domain satisfies the EoS $\rho=p=0$ and can be expressed by
the Reissner-Nordstr\"{o}m metric as
\begin{equation}\label{39}
ds^{2}_{+}=\left(1-\frac{2\mathcal{M}}{r}+\frac{\mathcal{Q}^{2}}
{r^{2}}\right)dt^{2}-\left(1-\frac{2\mathcal{M}}{r}+\frac{\mathcal{Q}^{2}}
{r^{2}}\right)^{-1}dr^{2}-r^{2}\left(d\theta^{2}+\sin^{2}\theta
d\phi^{2}\right),
\end{equation}
where $\mathcal{M}$ and $\mathcal{Q}$ are the total gravitational
mass and total charge, respectively. In the study of compact
objects, it is observed that there should be a smooth matching
between interior and exterior domains. Here, for a smooth connection
between $\mathcal{D}_{1}$ and $\mathcal{D}_{3}$, we assume a
formalism proposed by Israel (1966). At the joining surface, i.e.,
at $r=\mathcal{R}$, the coefficients of the metric
\begin{equation}\label{40}
ds^{2}=\mathcal{H}(r)dt^{2}-\frac{dr^{2}}{\mathcal{H}(r)}-r^{2}
\left(d\theta^{2}+\sin^{2}\theta d\phi^{2}\right),
\end{equation}
are continuous but their derivatives may not exist at this surface.
Nevertheless, it is possible to acquire the surface stress-energy
($\mathcal{S}_{\gamma\delta}$) with the assistance of Israel
formalism. Considering Lanczos equation, the expression for
$\mathcal{S}_{\gamma\eta}$ is illustrated by
\begin{equation}\label{41}
\mathcal{S}^{\gamma}_{\eta}=\frac{1}{8\pi}\left(\delta^{\gamma}_{\eta}
\tau^{\alpha}_{\alpha}-\tau^{\gamma}_{\eta}\right),
\end{equation}
where $\tau_{\gamma\eta}=\mathcal{K}^{+}_{\gamma\eta}
-\mathcal{K}^{-}_{\gamma\eta}$ leads to the discontinuity in the
surface of extrinsic curvature or second fundamental form. Here, the
signs $-$ and $+$ show the interior and exterior domains,
respectively. The extrinsic curvature at hypersurface ($\Pi$)
corresponding to two regions of the thin-shell is of the form
\begin{equation}\label{42}
\mathcal{K}^{\pm}_{\gamma\eta}=-\Upsilon^{\pm}_{\mu}\left[\frac{\partial^{2}
x_{\mu}}{\partial\Im^{\gamma}\partial\Im^{\eta}}+\Gamma^{\mu}_{\alpha\nu}
\frac{\partial x^{\alpha}\partial x^{\nu}}
{\partial\Im^{\gamma}\partial\Im^{\eta}}\right],
\end{equation}
where $\Im^{\gamma}$ symbolizes the intrinsic shell coordinates,
$\Upsilon^{\pm}_{\mu}$ indicates the unit normals at $\Pi$ defined
as
\begin{equation}\label{43}
\Upsilon^{\pm}_{\mu}=\pm\left|g^{\alpha\nu}\frac{\partial\mathcal{H}
}{\partial x^{\alpha}}\frac{\partial\mathcal{H} }{\partial
x^{\nu}}\right|^{-\frac{1} {2}}\frac{\partial\mathcal{H}}{\partial
x^{\mu}},\quad \textrm{with} \quad\Upsilon^{\mu}\Upsilon_{\mu}=1.
\end{equation}

Employing the Lanczos equations, the mathematical descriptions of
surface energy density ($\varrho$) and surface pressure
($\mathcal{P}$) in $\mathcal{S}_{\gamma\eta}=diag(\varrho,$
$-\mathcal{P},-\mathcal{P},-\mathcal{P})$, are given by
\begin{equation}\label{44}
\varrho=-\frac{1}{4\pi\mathcal{R}}
\left[\sqrt{\mathcal{H}}\right]^{+}_{-}, \quad
\mathcal{P}=\frac{-\varrho}{2}+\frac{1}{16\pi}
\left[\frac{\mathcal{H}'}{\sqrt{\mathcal{H}}}\right]^{+}_{-}.
\end{equation}
Substituting the considered $\mathcal{D}_{1}$ and $\mathcal{D}_{3}$
metrics along with the conformal vector in the above expressions, we
obtain
\begin{eqnarray}\label{45}
\varrho&=&-\frac{1}{4\pi\mathcal{R}}\left[\sqrt{1-\frac{2\mathcal{M}}
{\mathcal{R}}+\frac{\mathcal{Q}^{2}}{\mathcal{R}^{2}}}
-\tilde{\varphi}_{0}\mathcal{R}\right],\\\label{46}
\mathcal{P}&=&\frac{1}{8\pi\mathcal{R}}\left[\frac{1-\frac{\mathcal{M}}
{\mathcal{R}}}{\sqrt{1-\frac{2\mathcal{M}}{\mathcal{R}}
+\frac{\mathcal{Q}^{2}}{\mathcal{R}^{2}}}}-2
\tilde{\varphi}_{0}\mathcal{R}\right].
\end{eqnarray}
Using Eq.(\ref{45}), we evaluate the surface mass of the charged
gravastar thin-shell as
\begin{equation}\label{47}
\mathcal{M}_{shell}=4\pi\mathcal{R}^{2}\varrho=\mathcal{R}
\left[\tilde{\varphi}_{0}\mathcal{R}-\sqrt{1-\frac{2\mathcal{M}}
{\mathcal{R}}+\frac{\mathcal{Q}^{2}}{\mathcal{R}^{2}}}\right].
\end{equation}
This leads to the total mass of charged gravastar in the following
form
\begin{equation}\label{48}
\mathcal{M}=\frac{1}{2\mathcal{R}}\left[\mathcal{R}^{2}
+\mathcal{Q}^{2}-\mathcal{M}_{shell}^{2}\mathcal{R}^{2}+2
\tilde{\varphi}_{0}\mathcal{M}_{shell}\mathcal{R}^{4}
-\tilde{\varphi}_{0}^{2}\mathcal{R}^{6}\right].
\end{equation}

\section{Physical Aspects of Charged Gravastar}

In this section, we discuss some essential physical aspects such as
EoS parameter, entropy, length and the energy-thickness
relationship, that entirely portray the intermediate geometry of
charged gravastar.

\subsection{The EoS Parameter}

The EoS parameter at $r=\mathcal{R}$, using Eqs.(\ref{45}) and
(\ref{46}), turns out to be
\begin{equation}\label{49}
\omega(\mathcal{R})=\frac{\mathcal{P}}{\varrho}=\frac{1
-\frac{\mathcal{M}}{\mathcal{R}}-2\tilde{\varphi}_{0}\mathcal{R}
\sqrt{1-\frac{2\mathcal{M}}{\mathcal{R}}+\frac{\mathcal{Q}^{2}}
{\mathcal{R}^{2}}}}{2\left((1-\frac{2\mathcal{M}}{\mathcal{R}}
+\frac{\mathcal{Q}^{2}}{\mathcal{R}^{2}})-\tilde{\varphi}_{0}
\mathcal{R}\sqrt{1-\frac{2\mathcal{M}}{\mathcal{R}}
+\frac{\mathcal{Q}^{2}}{\mathcal{R}^{2}}}\right)}.
\end{equation}
This equation is comprised of different factors involving the
fractions with square root terms that enhance sensitivity of the
equation. In order to have real EoS parameter, the necessary
required condition is either $\frac{2\mathcal{M}}{\mathcal{R}}
-\frac{\mathcal{Q}^{2}}{\mathcal{R}^{2}}<1$ or
$\frac{2\mathcal{M}}{\mathcal{R}}$. These inequalities provide the
relation between $\mathcal{M}$, $\mathcal{R}$ and $\mathcal{Q}$ as
$\mathcal{Q}>\sqrt{\mathcal{R}(2\mathcal{M}-\mathcal{R})}$ and
$\mathcal{M}<\frac{\mathcal{R}}{2}$, respectively. The specification
of positive density as well as positive pressure always leads to the
positive value of $\omega$. For appropriately large value of
$\mathcal{R}$, we attain $\omega(\mathcal{R})\approx1$ (Usmani et
al. 2011). This large value of $\mathcal{R}$ manifests the structure
of gravastar like compact objects. Also, the insertion of some
particular value of $\mathcal{R}$ in Eq.(\ref{46}) may provide
$\mathcal{P}=0$, which corresponds to a dust shell.

\subsection{Length of the Shell}

We have considered that the geometry of the stiff fluid charged
gravastar shell is placed at $r=\mathcal{R}$ which is illustrated by
the domain $\mathcal{D}_{1}$. As the thickness of the intermediate
shell is supposed to be extremely small, i.e., $\epsilon\ll1$,
therefore, the domain $\mathcal{D}_{3}$ appears from the interface
$r=\mathcal{R}+\epsilon$. Thus the proper length ($\mathcal{L}$)
between two joints, i.e., the thin-shell is evaluated as
\begin{equation}\label{50}
\mathcal{L}=\int^{\mathcal{R}+\epsilon}_{\mathcal{R}}\sqrt{e^{\chi}}dr
=\sqrt{2}\left[r\sqrt{1-\frac{2\tilde{\varphi}_{1}}{r}}+\tilde{\varphi}_{1}
\ln(r+r\sqrt{1-\frac{2\tilde{\varphi}_{1}}{r}}
-\tilde{\varphi}_{1})\right]^{\mathcal{R}+\epsilon}_{\mathcal{R}}.
\end{equation}
The relation between the proper length and thickness of the charged
gravastar shell is exhibited in Figure \textbf{3} which interprets
the increasing as well as direct connection between them.
\begin{figure}\center
\epsfig{file=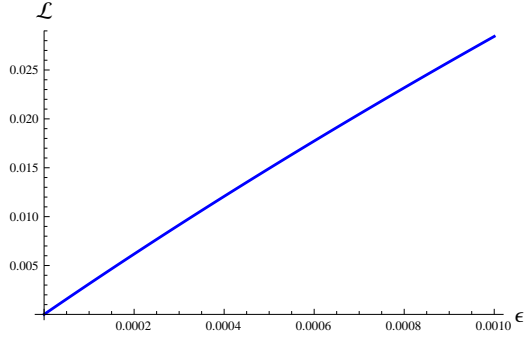,width=0.5\linewidth}\\
\caption{Interpretation of the proper length $(km)$ versus thickness
$(km)$ of the shell of charged gravastar with
$\tilde{\varphi}_{1}<\frac{r}{2}$.}
\end{figure}

\subsection{Entropy Inside the Shell}

In the study of gravastar, Mazur and Mottola (2004) discussed the
zero entropy density ($\mathcal{S}$) inside the domain
$\mathcal{D}_{1}$ which displays the consistency with a single
condensate region. In this work, the entropy inside the charged
gravastar shell is determined by (Das 2017)
\begin{equation}\label{51}
\mathcal{S}=\int^{\mathcal{R}+\epsilon}_{\mathcal{R}}4\pi
r^{2}\mathcal{U}(r)\sqrt{e^{\chi}}dr,
\end{equation}
where $\mathcal{U}(r)$ indicates the entropy density and
corresponding to radially dependent temperature ($\mathcal{T}(r)$),
it is expressed as
\begin{equation}\label{52}
\mathcal{U}(r)=\frac{\sigma^{2}K^{2}_{\mathcal{B}}\mathcal{T}(r)}{4\pi
\hbar^{2}}=\sigma\left(\frac{K_{\mathcal{B}}}{\hbar}\right)
\sqrt{\frac{p}{2\pi}}.
\end{equation}
Here $\sigma$ is a dimensionless constant and in this work, we take
Planckian units as $\hbar=K_{\mathcal{B}}=1$. Thus Eq.(\ref{51})
becomes
\begin{equation}\label{53}
\mathcal{S}=\int^{\mathcal{R}+\epsilon}_{\mathcal{R}}\frac{4\pi\sigma
r^{2}}{\sqrt{2\pi}}\sqrt{\frac{1}{2r^{2}(1+2\beta)}\left(1-\frac{3
\tilde{\varphi}_{1}}{r}\right)\left(\frac{1}{\frac{1}{2}
-\frac{\tilde{\varphi}_{1}}{r}}\right)}dr.
\end{equation}
Integration of the above equation yields
\begin{eqnarray}\nonumber
\mathcal{S}&=&\Bigg[\frac{\sigma
r\sqrt{\frac{\pi}{2}\left(\frac{r-3\tilde{\varphi}_{1}}{r^{2}(1+2\beta)
(r-2\tilde{\varphi}_{1})}\right)}}{\sqrt{r-3\tilde{\varphi}_{1}}}
\left\{\sqrt{r-3\tilde{\varphi}_{1}}(2r^{2}-r\tilde{\varphi}_{1}
-6\tilde{\varphi}_{1}^{2})\right.\Bigg.\\\label{54}&-&\left.\Bigg.9
\tilde{\varphi}_{1}^{2}(r-2\tilde{\varphi}_{1})\ln\left(\sqrt{r-2
\tilde{\varphi}_{1}}+\sqrt{r-3\tilde{\varphi}_{1}}\right)\right\}
\Bigg]^{\mathcal{R}+\epsilon}_{\mathcal{R}}.
\end{eqnarray}
On setting $\beta=0$, this reduces to the entropy equation of GR
along with conformal vector. The direct relation between entropy and
thickness of the charged gravastar shell for two positive values of
$\beta$ can be observed from Figure \textbf{4}.
\begin{figure}\center
\epsfig{file=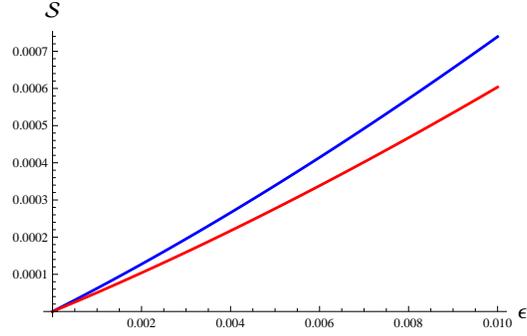,width=0.5\linewidth}\\
\caption{Variation of entropy inside the shell versus thickness of
the shell of charged gravastar with
$\tilde{\varphi}_{1}<\frac{r}{2}$, $\beta=\frac{1}{2}$ (blue) and
$\beta=1$ (red).}
\end{figure}

\subsection{Energy Inside the Shell}

In the domain $\mathcal{D}_{1}$, we have considered the EoS
$\rho=-p$ which reveals negative energy domain ensuring the
existence of repulsive nature of the domain $\mathcal{D}_{1}$. But
inside the shell of charged gravastar in $f(R,T)$ gravity, the
energy takes the form
\begin{equation}\label{55}
\mathcal{E}=\frac{1}{2}\int^{\mathcal{R}+\epsilon}_{\mathcal{R}}
\left(\rho+2\pi E^{2}\right)r^{2}dr=\frac{1}{2}\int^{\mathcal{R}
+\epsilon}_{\mathcal{R}}\left[\frac{1}{4(1+2\beta)}\left(1-\frac{3
\tilde{\varphi}_{1}}{r}\right)+\frac{3\tilde{\varphi}_{1}}{4r}\right]r^{2}dr,
\end{equation}
whose integration provides
\begin{equation}\label{56}
\mathcal{E}=\frac{\epsilon+6\beta\tilde{\varphi}_{1}[\ln(\mathcal{R}
+\epsilon)-\ln(\mathcal{R})]}{4(1+2\beta)}.
\end{equation}
This shows a direct relation of the energy with the thickness of the
shell. The graphical analysis of energy-thickness relation
corresponding to different values of $\beta$ is given in Figure
\textbf{5} which displays the non-repulsive nature of energy inside
the shell. This reduces to GR for $\beta=0$ (Usmani et al. 2011).
\begin{figure}\center
\epsfig{file=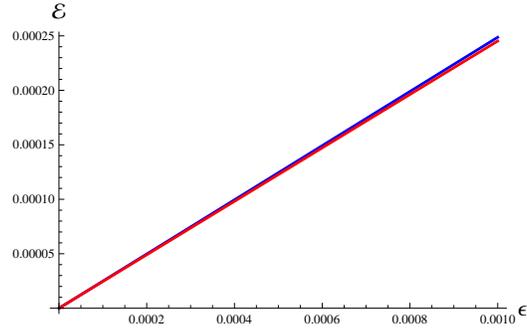,width=0.5\linewidth}\\
\caption{Variation of energy inside the shell versus thickness of
the shell with $\tilde{\varphi}_{1}<\frac{r}{2}$,
$\beta=\frac{1}{2}$ (blue) and $\beta=1$ (red).}
\end{figure}

\section{Concluding Remarks}

This paper is dedicated to discuss a novel unique stellar object in
$f(R,T)$ gravity admitting conformal Killing vectors which is
originally proposed in GR for gravastar (Mazur and Mottola 2004). We
have considered an internally charged gravastar with the exterior
expressed by the Reissner-Nordstr\"{o}m metric. The $f(R,T)=R+2\beta
T$ gravity model is used to study the geometry of charged gravastar
for three domains along with distinct EoS. We have evaluated exact
non-singular solutions of the collapsing system in terms of
conformal vector. These solutions yield some fascinating aspects of
the charged gravastar which are physically consistent in this
gravity.

In order to analyze the structure of charged gravastar, we have
considered that the interior domain $\mathcal{D}_{1}$ satisfies the
EoS $\rho=-p$, which shows the repulsive pressure describing the
dark energy. We have studied the behavior of gravitational mass in
$\mathcal{D}_{1}$ with respect to the radius $r_{1}=\mathcal{R}$.
The domain $\mathcal{D}_{1}$ is bounded by a thin-shell which is
composed of ultra-relativistic stiff fluid and obeys the EoS
$\rho=p$, indicating the non-repulsive nature of charged gravastar
shell. The exterior domain $\mathcal{D}_{3}$ is assumed to be
entirely vacuum executing the EoS $\rho=p=0$. We have found
constraints on conformal vector $\varphi$ as well as model parameter
$\beta$ from the resulting exact solutions in these domains. The
graphical behavior of charge and energy density of stiff fluid are
shown for positive values of $\beta$ against the thickness of the
shell.

We have studied some interesting physical features of charged
gravastar such as length, energy, entropy and EoS parameter. For the
real value of EoS parameter, we must have $\frac{2\mathcal{M}}
{\mathcal{R}}+\frac{\mathcal{Q}^{2}}{\mathcal{R}^{2}}<1$, which may
provide some specific value of $\omega$ that could result in the
stability of charged gravastar. We have also found that only for
large appropriate value of $\mathcal{R}$, we obtain
$\omega(\mathcal{R})\approx1$ which provides the EoS $p=\rho$ in the
intermediate domain. We have figured out the expressions of length,
energy as well as entropy and analyzed graphically their direct
relations with the thickness of the charged gravastar shell for
positive values of $\beta$.

Usmani et al. (2011) discussed the geometry of charged gravastars
with conformal Killing vectors in GR but they did not analyze their
solutions graphically. Das et al. (2017) found exact solutions of
gravastars in $f(R,T)$ scenario without charge as well as conformal
vector and presented the graphical analysis of density, pressure,
length, energy and entropy inside the shell. It is worthwhile to
mention here that we have introduced the conformal motion as well as
effects of charge in $f(R,T)$ framework to describe the geometry of
gravastars and also shown the graphical behavior of our obtained
solutions. We have found that the gravitational mass obtained from
the matching of interior and Reissner-Nordstr\"{o}m exterior metrics
in this gravity becomes an electromagnetic mass that provides an
attractive force to overcome the repulsive effects of charge which
is consistent with GR (Usmani et al. 2011). These repulsive effects
also counterbalance the attractive force of gravity and lead to the
more stable configuration of gravastar than the structure defined
without charge (Das et al. 2017). Moreover, solutions for dark
energy EoS demonstrate that charged gravastars are connected with
dark stars. We would like to mention here that all the obtained
solutions reduce to GR for $\beta=0$ (Usmani et al. 2011).

\vspace{0.5cm}

{\bf Acknowledgment}

\vspace{0.25cm}

We would like to thank the Higher Education Commission, Islamabad,
Pakistan for its financial support through the {\it Indigenous Ph.D.
5000 Fellowship Program Phase-II, Batch-III.}

\end{document}